\documentclass[sigconf,natbib,authorversion]{acmart}

\usepackage[english]{babel}
\usepackage[inline]{enumitem}
\usepackage{numprint}

\npdecimalsign{.}

\copyrightyear{2023}
\acmYear{2023}
\setcopyright{acmlicensed}\acmConference[SIGIR '23]{Proceedings of the 46th International ACM SIGIR Conference on Research and Development in Information Retrieval}{July 23--27, 2023}{Taipei, Taiwan}
\acmBooktitle{Proceedings of the 46th International ACM SIGIR Conference on Research and Development in Information Retrieval (SIGIR '23), July 23--27, 2023, Taipei, Taiwan}
\acmPrice{15.00}
\acmDOI{10.1145/3539618.3591849}
\acmISBN{978-1-4503-9408-6/23/07}

\begin{document}

\title{Modeling Spoken Information Queries for Virtual Assistants}
\subtitle{Open Problems, Challenges and Opportunities}

\author{Christophe Van Gysel}
\email{cvangysel@apple.com}
\affiliation{%
  \institution{Apple}
  \city{Cambridge, MA}
  \country{USA}
}


\newcommand{\IRRelevance}[1]{\noindent\textbf{Relevance to IR.} #1}
\renewcommand{\shortauthors}{Van Gysel}

\begin{abstract}
Virtual assistants are becoming increasingly important speech-driven Information Retrieval platforms that assist users with various tasks.
We discuss open problems and challenges with respect to modeling spoken information queries for virtual assistants, and list opportunities where Information Retrieval methods and research can be applied to improve the quality of virtual assistant speech recognition.
We discuss how query domain classification, knowledge graphs and user interaction data, and query personalization can be helpful to improve the accurate recognition of spoken information domain queries. %
Finally, we also provide a brief overview of current problems and challenges in speech recognition.
\end{abstract}


\maketitle

\section{Introduction}

Virtual assistants (VAs) are becoming increasingly important \cite{Juniper2019popularity} Information Retrieval (IR) platforms that assist users with various tasks. Users primarily interact with VAs through voice commands, where users initiate a retrieval request by uttering a query, possibly preceded by a wake word (e.g., \emph{``hey VA''}).
Accurately transcribing spoken voice queries \cite{Guy2016searching}, for subsequent processing by the retrieval engine, is a challenging problem that can benefit greatly from knowledge of the IR application.

Automated Speech Recognition (ASR) systems, responsible for transcribing the spoken utterance, are trained on audio/text pairs that are expensive to obtain. Language models (LMs) are a component within ASR systems that act as a query prior and are trained on only text. The LM becomes increasingly important when the spoken query is ambiguous or difficult to understand (e.g., unintelligible speech).
Take as an example the encyclopedia query \emph{``what is borrelia''} where the user intends to obtain information about the Borrelia bacteria. In this particular case, the entity \emph{borrelia} may be misrecognized as \emph{gorilla} if the LM assigns a low likelihood to the conditional probability $\text{P}\left( \emph{borrelia} \mid \text{what is} \right)$. However, this problem can be alleviated through query log analysis, injection of external knowledge (e.g., entity popularity), and use of contextual signals, amongst other methods.

The challenging nature of VA query recognition is further exacerbated by stringent runtime requirements. Recognition needs to occur in real-time as users expect results soon after they finish speaking. When ASR occurs on-device, model size becomes an additional constraint---since disk space and network bandwidth are costly. In addition, the ability to patch or perform incremental updates of models is desirable functionality. Finally, LMs need to be trained within a reasonable amount of time, since otherwise they may be outdated by the time the LMs reach edge devices.

We provide a succinct overview of the use of LMs in ASR, and subsequently cover topics on using knowledge of the IR application to improve ASR with a focus on entity-heavy VA queries: %
\begin{enumerate*}[label=(\arabic*)]
  \item query domain classification,
  \item entity popularity and knowledge graph (KG) mining, and
  \item personalization.
\end{enumerate*}
Finally, we briefly cover non-IR topics relating to LMs and ASR.

\section{ASR Primer}

\newcommand{\StartToken}{\text{<s>}}
\newcommand{\EndToken}{\text{</s>}}

\newcommand{\Apply}[2]{#1\left(#2\right)}
\newcommand{\Prob}[2][P]{\Apply{#1}{#2}}
\newcommand{\CondProb}[3][P]{\Prob[#1]{#2 \mid #3}}

Automated Speech Recognition is the task of translating a speech signal $X$ into a string of words $s$. %
Contemporary ASR systems can be divided into two categories: %
\begin{enumerate*}[label=(\alph*)]
  \item traditional \textbf{hybrid systems} that rely on Bayes' rule to combine acoustic and language model components, and
  \item more modern \textbf{end-to-end systems} that directly predict output sequences of text from acoustic representations.
\end{enumerate*}

\subsection{Hybrid ASR systems}

Hybrid ASR systems operate by decomposing the ASR task using Bayes' rule as follows \citep[p.~289]{Jurafsky2008slp}:
\begin{equation}
\begin{split}
S^* & = \text{argmax}_s \CondProb{s}{X} = \text{argmax}_s \frac{\CondProb{X}{s}}{\Prob{X}} \cdot \Prob{s} \\
    & = \text{argmax}_s \CondProb{X}{s} \cdot \Prob{s}, \\
\end{split}
\end{equation}
where $\CondProb{X}{s}$ is provided by the acoustic model (AM) and denotes the likelihood of speech signal $X$ given the string of words $s$, and $\Prob{s}$ is provided by the LM and denotes the prior probability of a string of words. $\Prob{X}$ is the probability of the speech signal and can be ignored as it is constant for all hypotheses. The AM and LM are trained independently and subsequently combined.

\subsection{End-to-end ASR systems}

End-to-end ASR (E2E) systems directly compute the probability distribution $\CondProb{s}{X}$ of output strings of words $s$ given speech signals $X$, and are typically implemented using neural encoder-decoder architectures \citep[\S16.3]{Jurafsky2023slp3draft}. Since E2E systems model the ASR task directly, they are trained on paired audio-text data--which can be expensive to obtain and may not provide full coverage for tail utterances.
Hence, often, an additional LM, trained only on abundantly available texts, and external to the E2E model, is combined through interpolation as follows:
\begin{equation}
\begin{split}
S^* & = \text{argmax}_s \CondProb{s}{X} \approx \text{argmax}_s \CondProb[P_\text{E2E}]{s}{X} \cdot \Prob[P_\text{Ext. LM}]{s}^\lambda, \\
\end{split}
\label{eq:e2e}
\end{equation}
where $\CondProb[P_\text{E2E}]{s}{X}$ is provided by the E2E model, $\Prob[P_\text{Ext. LM}]{s}$ is the LM probability and $\lambda$ is an interpolation hyperparameter.

\subsection{Language models}

Regardless of ASR system architecture, hybrid or E2E, a LM trained solely on text data can be used to improve recognition quality.
The LM builds on the chain rule of probability:
\begin{equation}
    \Prob{W} = \Prob{w_1 w_2\dots w_n} = \prod^{N}_{i=1}\CondProb{w_i}{w_1 w_2\dots w_{i-1}}.
\end{equation}
In practice, strings of words are wrapped in special start/end markers ($\StartToken{}$/$\EndToken{}$, resp.) to denote the beginning and the end of the string of words (which is, typically, a sentence). For example, the prior probability of utterance \emph{SIGIR} would be computed as
\begin{equation*}
\Prob{\text{<s> SIGIR </s>}} = \CondProb{\text{SIGIR}}{\StartToken{}} \CondProb{\EndToken{}}{\text{<s> SIGIR}},
\end{equation*}
where <s> and </s> mark the beginning and end of sentence, resp.

\section{Open Problems and Challenges}

\subsection{Use of query domain classifications}
\label{sec:domain}

VA queries can be categorized according to domains where each domain supports specific use-cases. For example, there exist media player queries such as \emph{``play the look by metronomy''} where the user instructs the VA to play the song ``The Look'' by the band Metronomy, or encyclopedic queries where the user wants to learn more about a specific entity (e.g., \emph{``who is joe biden''}).

In this section we discuss the application of query domain classifications, possibly provided by NLP/IR methods, to improve VA ASR, either by %
\begin{enumerate*}[label=(\alph*)]
  \item using domain classifications at runtime to guide the ASR decoding process, or
  \item utilizing the classification of queries at LM training time.
\end{enumerate*}

\subsubsection{Improving the ASR decoding process}

At recognition time, contextual signals, such as partial recognition hypotheses \citep{Pusateri2019interpolation} or the user location \citep{Xiaoqiang2018geographic}, can be used to modify the search space. \citet{Pusateri2019interpolation} combine multiple domain-specific expert n-gram LMs into a single LM by weighing the expert LMs based on the confidence expressed by each expert LM on how well they support specific left spoken contexts. Following the example above, the media player domain LM would receive a large weight following the left context \emph{``<s> play''}, whereas a LM trained on encyclopedia queries may be well-suited following left context \emph{``<s> who is''}.

\IRRelevance{From an IR perspective, contextual signals extracted from (partial) user interactions (e.g., session information, partial queries) with the VA can be integrated into the ASR component responsible for combining multiple domain-specific expert models. Effective integration of contextual signals into E2E ASR systems (i.e., Eq.~\ref{eq:e2e}) remains an open problem today.}

\subsubsection{Building better LMs by leveraging query domain classifications}

\citet{Gondala2021error} take a different approach and use classifications of training data queries to influence the n-gram LM training algorithm. For example, query domains that reference many tail entities can be allocated more model capacity and that in turn improves the recognition of tail entities.

\IRRelevance{%
Offline classification of query logs can be used to improve ASR. While in \citep{Gondala2021error}, the authors used a domain-driven generative process \citep{VanGysel2022phirtn} to obtain training query texts, their approach can also be applied on queries that occur in usage logs.
However, ASR is a noisy process and consequently queries may contain recognition errors, and hence, domain classification methods designed for typed query traffic may not directly apply.
The IR community may find the usage of signals made available during the ASR decoding process, such as word-level confidence \citep{Jeon2020confidence}, helpful to adapt methods designed for typed query classification to spoken queries.}

\subsection{KGs and other external data}

As mentioned at the end of the previous section, spoken query logs contain recognition errors, and LMs used for ASR are often trained, at least partially, on query logs. This practice can lead to a feedback loop of reinforced errors. While filtering techniques can provide some relief, they may also introduce undesirable biases in the training data. The use of external data sources is an alternative solution that can benefit the recognition of entity-rich queries.

\subsubsection{Query templatization and entity popularity}

\citet{Gandhe2018lmadaptation} estimate n-gram LMs directly from entity-rich grammars to improve ASR for new application intents in VAs. In this case, queries such as in the example of \S\ref{sec:domain} can be represented as templates (e.g., \emph{``play \$SONG by \$ARTIST''}) with entity slots. \citet{VanGysel2022phirtn} released a VA media player query grammar, including a large list of media player entities extracted from a large-scale media catalog user interactions. In \citep{VanGysel2020recency}, the authors extract entities from a VA query log that occur in the presence of spoken left context (e.g., a verb) to improve the recognition of entity name queries \citep{Yin2010nameentityqueries} in the absence of left context.

\IRRelevance{From an IR point of view, there exist multiple challenges. First of all, while query templates can be created manually by domain experts, methods to automatically extract templates from a query log can be useful. Secondly, while entity popularity can be extracted from external sources, there likely still exists a gap between popularity in the source application and the VA application (e.g., a difference in demographics). Hence, entity popularity adaptation methods are still an open research problem.}

\subsubsection{Using KG relations during ASR decoding}

\citet{Saebi2021entityaware}, amongst others \citep{Logan2019factaware,Hayashi2020latent}, make use of entity type and entity--entity relations during the ASR decoding process to improve the recognition of tail named entities. For example, if the ASR decoder is considering two hypotheses %
\begin{enumerate*}[label=(\alph*)]
  \item \emph{``play can you moon by Harry Styles''} and
  \item \emph{``play Canyon Moon by Harry Styles''},
\end{enumerate*}
their approach will use the KG relationship between artist (i.e., ``Harry Styles'') and song title (i.e., ``Canyon Moon'') as a signal during recognition to boost the likelihood that the factually correct hypothesis (b) is chosen.

\IRRelevance{Hence, IR research focusing on improving KGs and entity linking in spoken queries can directly improve the effectiveness of VA ASR.}

\subsection{Personalization}

Personalization of on-device ASR is an active area of research \cite{Breiner2022userlibri}. For the VA application, and from the language modeling perspective (as opposed to acoustics \citep{Sim2019personalizationinvestigation}), systems may be able to benefit from signals used in other search applications, such as Web search \cite{Sieg2007webpersonalization}, as users with different profiles tend to search for different sets of topics. More specifically, knowledge about the user's interests--which may eventually lead to an interaction with a specific intent--can be helpful to improve user experience.
\citet{Xiaoqiang2018geographic} improve ASR by bucketing users according to their coarse geographic location and enable region-specific query LMs during the ASR decoding process. By personalizing the ASR query model based on user location, they show a significant improvement in the accurate recognition of spoken point-of-interest queries.

\IRRelevance{On the IR side, user models \cite{Chuklin2015clickmodels} based on query behavior or other signals may be helpful to power futher personalization of on-device VA ASR.}

\subsection{Beyond IR}

In the previous sections, we focused on the impact of IR research on the accurate recognition of spoken information queries for the VA application. Naturally, there exist a multitude of challenges on the ASR side as well. %
End-to-end ASR models \citep{Wang2019overview}, as opposed to traditional Gaussian mixture models, have been increasingly gaining popularity since end-to-end models consist of less components---hence, reducing maintenance costs. However, integration of external LMs into \citep{Shan2019component,Cabrera2021lmfusion,Kim2021lmfusion}, and personalization of \citep{Sim2019personalizationinvestigation,Sim2019personalization,Gourav2021personalization}, end-to-end systems remains an active research area. %
With respect to LM, Neural Network LMs (NNLM) \citep{Bengio2000nnlm} have gained popularity within ASR \citep{Graves2013hybrid,Zhang2015fofe,Shewalkar2019performance}. In the case of VA ASR, NNLMs can be significantly more economical in terms of storage costs than their N-Gram LM \citep{Katz1987backoff} counterparts, as with the latter, the size of the models grows proportional to the data complexity.
However, practical limitations such as training an NNLM from a heterogenous corpora, inference latency \citep{Raju2019scalable}, and federated learning \citep{Xu2022federated} remain challenging.
More recently, large pre-trained Transformer LMs \citep{Vaswani2017attention,Devlin2019bert,Brown2020gpt3,OpenAI2023gpt4} have also been used to improve ASR \citep{Shin2019bertasr,Huang2019transformers,Chiu2021innovative,Xu2022rescorebert}, although a domain gap may exist \citep{Pelloin2022asr}; and have also been shown to be effective for synthetic data generation \citep{Bonifacio2022inpars,Hamalainen2023evaluating}.

\section{Conclusions}

We discussed open problems and challenges with respect to modeling spoken information queries for VAs, and listed opportunities where IR methods and research can be applied to improve the quality of VA ASR.
More specifically, we discussed how query domain classification can be used during speech recognition and to build better LMs. Next, we discussed the use of KGs and external data sources based on user interactions, and discussed personalization. Finally, and for completeness, we briefly provided an overview of challenges and open problems within ASR.

We hope that the discussed topics are useful to IR researchers and lead to the exploration of new, cross-disciplinary research directions and ideas, and as inspiration to discover new application domains for existing methods.

\section*{Acknowledgments}

The authors would like to thank %
Manos Tsagkias, %
Ernest Pusateri, %
Barry Theobald, %
Man-Hung Siu, %
Ilya Oparin, %
and the anonymous reviewers %
for their comments and feedback.

\section*{Speaker biography}

Christophe Van Gysel is a Staff Research Scientist working on the Siri Speech language modeling team at Apple where he works on the boundary between ASR and Search.
Christophe obtained his PhD in Computer Science from the University of Amsterdam in 2017. During his PhD, Christophe worked on neural ranking using representation learning models with a focus on entities and published at WWW, SIGIR, CIKM, WSDM, TOIS, amonst others.

\section*{Company profile}

Apple revolutionised personal technology with the introduction of the Macintosh in 1984. Today, Apple leads the world in innovation with iPhone, iPad, Mac, Apple Watch, and Apple TV. Apple’s five software platforms — iOS, iPadOS, macOS, watchOS, and tvOS — provide seamless experiences across all Apple devices and empower people with breakthrough services including the App Store, Apple Music, Apple Pay, and iCloud. Apple’s more than 100,000 employees are dedicated to making the best products on earth, and to leaving the world better than we found it.

\bibliographystyle{ACM-Reference-Format}
\bibliography{sigir2023-query-modeling-industry}

\end{document}